\documentclass[trackchanges]{aastex701}
\usepackage{graphicx,times}             
\usepackage{natbib}
\usepackage{amssymb,amsmath}
\usepackage{lipsum}
\usepackage{threeparttable}
\usepackage{gensymb}  
\usepackage{siunitx}  
\usepackage{placeins} 
\usepackage[utf8]{inputenc}
\DeclareUnicodeCharacter{2212}{-}
\renewcommand{\textbf}[1]{#1}

\shorttitle{Stochastic variability and rms--flux in EP240309a}
\shortauthors{Wu et al.}

\begin{document}

\title{Stochastic Optical Variability and an rms––flux Relation in the Intermediate Polar EP240309a}

\correspondingauthor{Y.-D. Hu}
\email{huyoudong072@hotmail.com}

\author[0009-0004-7113-8258]{S.-Y. Wu}
\affiliation{Instituto de Astrof\'{\i}sica de Andaluc\'{\i}a (IAA-CSIC), Glorieta de la Astronom\'{\i}a s/n, 18080 Granada, Spain}
\affiliation{Department of Physics and Mathematics, University of Granada, 18012 Granada, Spain}
\email{wusiyu.11@outlook.com}

\author[0000-0002-7400-4608]{Y.-D. Hu}
\affiliation{Guangxi Key Laboratory for Relativistic Astrophysics, School of Physical Science and Technology, Guangxi University, Nanning 530004, China}
\email{huyoudong072@hotmail.com}

\author[0000-0002-7273-3671]{I.~P\'{e}rez-Garc\'{\i}a}
\affiliation{Instituto de Astrof\'{\i}sica de Andaluc\'{\i}a (IAA-CSIC), Glorieta de la Astronom\'{\i}a s/n, 18080 Granada, Spain}
\affiliation{Departamento de Ingenier\'ia de Sistemas y Autom\'atica, Unidad Asociada al CSIC por el IAA, Escuela de Ingenier\'ias Industriales, Universidad de M\'alaga, C.~Dr.~Ortiz Ramos s/n, 29071 M\'alaga, Spain}
\email{ipg@iaa.es}

\author[0000-0003-2999-3563]{A. J. Castro-Tirado}
\affiliation{Instituto de Astrof\'{\i}sica de Andaluc\'{\i}a (IAA-CSIC), Glorieta de la Astronom\'{\i}a s/n, 18080 Granada, Spain}
\affiliation{Departamento de Ingenier\'ia de Sistemas y Autom\'atica, Unidad Asociada al CSIC por el IAA, Escuela de Ingenier\'ias Industriales, Universidad de M\'alaga, C.~Dr.~Ortiz Ramos s/n, 29071 M\'alaga, Spain}
\email{ajct@iaa.es}

\author[0000-0003-4268-6277]{M. Gritsevich}
\affiliation{Instituto de Astrof\'{\i}sica de Andaluc\'{\i}a (IAA-CSIC), Glorieta de la Astronom\'{\i}a s/n, 18080 Granada, Spain}
\affiliation{Faculty of Science, University of Helsinki, Gustaf Hallstr\"omin katu 2, FI-00014 Helsinki, Finland}
\affiliation{Institute of Physics and Technology, Ural Federal University, Mira str.~19, 620002 Ekaterinburg, Russia}
\email{maria.gritsevich@helsinki.fi}

\author[0009-0009-4604-9639]{E.~J.~Fern\'{a}ndez-Garc\'{\i}a}
\affiliation{Instituto de Astrof\'{\i}sica de Andaluc\'{\i}a (IAA-CSIC), Glorieta de la Astronom\'{\i}a s/n, 18080 Granada, Spain}
\email{emifdez@iaa.es}

\author[0000-0001-7920-4564]{M.~D.~Caballero-Garc\'{\i}a}
\affiliation{Instituto de Astrof\'{\i}sica de Andaluc\'{\i}a (IAA-CSIC), Glorieta de la Astronom\'{\i}a s/n, 18080 Granada, Spain}
\email{mcaballero@iaa.es}

\author[0000-0003-2628-6468]{S. Guziy}
\affiliation{Instituto de Astrof\'{\i}sica de Andaluc\'{\i}a (IAA-CSIC), Glorieta de la Astronom\'{\i}a s/n, 18080 Granada, Spain}
\affiliation{Petro Mohyla Black Sea National University, Mykolaiv 54000, Ukraine}
\email{gssgrb@gmail.com}

\author[0000-0002-9899-2292]{G. Garc\'ia-Segura}
\affiliation{Instituto de Astronom\'ia, Universidad Nacional Aut\'onoma de M\'exico (IA-UNAM), Ensenada, Baja California, Mexico}
\affiliation{Instituto de Astrof\'{\i}sica de Andaluc\'{\i}a (IAA-CSIC), Glorieta de la Astronom\'{\i}a s/n, 18080 Granada, Spain}
\email{ggs@astro.unam.mx}

\author[0000-0002-7158-5099]{R. S\'{a}nchez-Ram\'{\i}rez}
\affiliation{Instituto de Astrof\'{\i}sica de Andaluc\'{\i}a (IAA-CSIC), Glorieta de la Astronom\'{\i}a s/n, 18080 Granada, Spain}
\email{ruben@iaa.es}

\author[0000-0002-5740-7747]{C. D. Kilpatrick}
\affiliation{Center for Interdisciplinary Exploration and Research in Astrophysics (CIERA), Northwestern University, Evanston, IL, USA}
\email{ckilpatrick@northwestern.edu}

\author[0000-0003-4383-2969]{C. R. Bom}
\affiliation{Centro Brasileiro de Pesquisas F\'isicas (CBPF), Rio de Janeiro, Brazil}
\email{debom@cbpf.br}

\author[0000-0003-3402-6164]{L. Santana}
\affiliation{Centro Brasileiro de Pesquisas F\'isicas (CBPF), Rio de Janeiro, Brazil}
\email{luidhysantana@gmail.com}

\author{A. Santos}
\affiliation{Centro Brasileiro de Pesquisas F\'isicas (CBPF), Rio de Janeiro, Brazil}
\email{andsouzasanttos@gmail.com}

\author[0000-0001-8890-5418]{P. J. Meintjes}
\affiliation{University of the Free State, Bloemfontein, South Africa}
\email{MeintjPJ@ufs.ac.za}

\author[0009-0004-9747-7215]{H. J. van Heerden}
\affiliation{University of the Free State, Bloemfontein, South Africa}
\email{VanHeerdenHJ@ufs.ac.za}

\author[0000-0001-5108-0627]{A. Mart\'in-Carrillo}
\affiliation{University College Dublin, Dublin, Ireland}
\email{antonio.martin-carrillo@ucd.ie}

\author[0000-0003-2931-3732]{L. Hanlon}
\affiliation{University College Dublin, Dublin, Ireland}
\email{lorraine.hanlon@ucd.ie}

\author{A. Maury}
\affiliation{San Pedro de Atacama Celestial Explorations, Chile}
\email{amaury@spaceobs.com}

\author[0000-0002-6809-9575]{D.-R. Xiong}
\affiliation{Yunnan Observatories, Chinese Academy of Sciences, 396 Yangfangwang, Guandu District, Kunming 650216, China}
\email{xiongdingrong@ynao.ac.cn}

\author[0000-0003-4111-5958]{B.-B. Zhang}
\affiliation{School of Astronomy and Space Science, Nanjing University, 163 Xianlin Avenue, Nanjing 210023, China}
\affiliation{Key Laboratory of Modern Astronomy and Astrophysics (Nanjing University), Ministry of Education, Nanjing 210023, China}
\email{bbzhang@nju.edu.cn}

\begin{abstract}
\textbf{Magnetic cataclysmic variables provide a natural laboratory for studying how accretion interacts with compact-object magnetospheres and generates stochastic variability. We present an optical variability study of the intermediate-polar candidate EP240309a, an \textit{Einstein Probe} X-ray transient, using BOOTES photometry, high-cadence TESS light curves, and a SOAR/Goodman optical spectrum. Previous studies found a white-dwarf spin period of 3.97\,min ($P_{\rm spin}\simeq238$\,s) and an orbital period of $P_{\rm orb}=3.7614(4)$\,h. Power spectral densities from the BOOTES data are consistent with single power laws with slopes $\alpha\simeq1.2$--1.8, with no statistically significant evidence for a bend across the sampled frequency range. Using red-noise simulations and injection--recovery tests, we place one-sided constraints on any putative break frequency, which translate, under standard dynamical identifications, into an upper limit on the magnetospheric radius of $R_{\rm m}\lesssim{\rm few}\times10^{10}$\,cm for $M_{\rm WD}=0.8\,M_\odot$. In the TESS data, we detect a linear rms--flux relation on hour timescales in three high-cadence sectors, while two other sectors do not show a robust detection, indicating epoch-dependent rms--flux behavior. The SOAR spectrum shows Balmer and He\,\textsc{ii} emission lines with ${\rm FWHM}\approx1000$--$1600\,{\rm km\,s^{-1}}$; under a Keplerian interpretation, these imply characteristic radii of $r\approx(0.9$--$3.4)\times10^{10}$\,cm, broadly comparable to the timing-based constraints. Overall, the data provide conservative, order-of-magnitude radius constraints consistent with accretion onto a magnetic white dwarf, but they do not establish the detailed accretion geometry or exclude stream-fed or mixed accretion scenarios.}
\end{abstract}




\section{Introduction}\label{sect:intro}

Magnetic cataclysmic variables (mCVs) are accreting white-dwarf binaries in which a megagauss-level magnetic field channels the flow and can truncate the inner disc. Intermediate polars (IPs) represent the disc-dominated end of the mCV family and are prolific X-ray emitters owing to accretion shocks and reprocessing \citep[e.g.,][]{patterson1984,warner1995,hellier1997}. Large-area surveys have expanded the IP census and enabled population studies from soft to hard X-rays \citep{Brunschweiger2009,Brunner2022}. In parallel, high-cadence optical monitoring shows that many cataclysmic variables (CVs) exhibit red-noise flickering and, in a growing number of cases, a linear rms--flux relation interpreted as the imprint of inward-propagating accretion-rate fluctuations in a stratified disc \citep[e.g.,][]{Uttley2001,uttley2005,Scaringi2012,Bruch2021,Bruch2022}.

The \textit{Einstein Probe} mission recently reported the transient EP~J115415.8$-$501810 (hereafter EP240309a), detected at $t_{\rm trig}=\mathrm{2024\ Mar\ 09\ 01{:}44{:}59\ UTC}$ (\citealt{ling2024}). Unless otherwise stated, times are referenced to $t_{\rm trig}$.
Follow-up studies have classified it as an IP candidate and established key coherent periods from long baselines: an orbital modulation near 3.762~h in \textit{TESS} photometry and a white-dwarf spin near 3.97~min in high-speed ground-based data \citep{potter2024,xiao2025,wu2025early}. These results secure the basic system parameters and motivate a deeper look at its stochastic optical variability and rms--flux/PSD behavior as complementary probes of the accretion flow.

In this work we combine multi-night BOOTES optical light curves, \textit{TESS} light curves, and a optical spectrum on the Southern Astrophysical Research (SOAR) Telescope to place EP240309a in the broader framework of accretion-driven variability. Our goals are threefold: (i) to establish and assess the robustness of a linear rms--flux relation on timescales of hours using space-based photometry, in which the rms variability amplitude is approximately proportional to the mean flux; (ii) to quantify the broadband power spectral density (PSD) of the longest ground-based runs and assess whether the data require a characteristic frequency (e.g., a bend or break) once narrow spin/beat/orbital components are accounted for; and (iii) to use deconvolved Balmer and He\,\textsc{ii} line widths to obtain order-of-magnitude constraints on characteristic radii in the disc and compare them with limits implied by the timing analysis. Taken together, these measurements connect phenomenological timing diagnostics with simple spectroscopic estimates of the dynamical scale in an IP. \textbf{The main purpose of the present work is therefore not to revisit the basic classification of EP240309a, but rather to provide a complementary characterization of its stochastic optical variability, to test whether the rms--flux behavior is persistent across multiple \textit{TESS} epochs, and to place correspondingly conservative constraints on the accretion geometry.}

\textbf{As a consistency check, we searched the BOOTES light curves for the published spin and orbital periods using Lomb--Scargle \citep{Lomb1976} and phase-dispersion minimization \citep{Stellingwerf1978}.
Owing to the few-hour nightly windows and strong red noise, no independent significant detections are obtained; the BOOTES data are therefore consistent with, but do not improve upon, the \textit{TESS}-based timing solution.
We thus adopt the published periods for narrow-line prewhitening prior to continuum PSD modeling.}

The paper is organized as follows. Section~\ref{sec:data} describes the BOOTES and SOAR observations and reductions, and summarizes the \textit{TESS} data products. Section~\ref{sec:methods} outlines the analysis workflow for prewhitening, PSD estimation and model comparisons, red-noise surrogate tests, and rms--flux measurements. Section~\ref{sec:results} presents the light curves, optical spectroscopic results, periodicity checks, PSD analysis, and the \textit{TESS} rms--flux relation. Section~\ref{sec:discussion} \textbf{interprets these findings in terms of stochastic accretion variability in magnetic cataclysmic variables and discusses conservative, order-of-magnitude radius constraints derived from timing and line-width diagnostics, emphasizing that they provide consistency checks rather than a proof of disc-fed accretion or magnetospheric truncation}. Section~\ref{sec:conclusions} summarizes our main conclusions.

\section{Observations}\label{sec:data}

We monitored EP240309a from 2024 March 21 (UTC 22:47) to March 25 (UTC 07:07)
in Clear (unfiltered) mode and report magnitudes calibrated onto an $R$-band scale
with a nominal 30\,s cadence (30\,s exposures for most runs; 180\,s exposures on the first night).
Observations were taken with two 0.6\,m Ritchey--Chr\'etien telescopes in the BOOTES network
\citep[see][for a recent overview]{CastroTirado2023NatAstron}:
BOOTES-6 (B6; Boyden Astronomical Observatory, South Africa) and BOOTES-7 (B7; San Pedro de Atacama, Chile).
Both use Andor iXon X3 electron-multiplying CCD (EMCCD)~888 cameras ($1024\times 1024$\,pix; $0.6^{\prime\prime}$/pix),
cooled to $-50.0\pm0.1^\circ$C.

Standard reduction was applied with custom Python routines based on the \texttt{astropy}, \texttt{numpy}, \texttt{scipy}, and \texttt{matplotlib} packages, following the standard BOOTES EMCCD procedures (e.g., \citealt{Castro1999,hu2023}): bias subtraction (median stacks of zero-second
frames), twilight-flat correction, and source detection on the reduced images.
Aperture photometry used a 5-pixel circular aperture with a 7--9 pixel background annulus.
Photometric zero points were tied to \textit{Gaia} Data Release~3 (DR3) field stars via a flux-weighted fit with $1.5\sigma$ outlier rejection \citep{GaiaDR3}.
Because the BOOTES frames were obtained in Clear (unfiltered) mode, we place the differential photometry onto an $R$-band magnitude scale through the same field-star calibration.
Uncertainties combine photon and background noise, CCD read noise, and the zero-point scatter; typical per-point uncertainties are $0.01$--$0.03$\,mag.
We obtained several BOOTES time-series runs between 2024 March 21--25 with cadences spanning approximately 50--335\,s depending on the exposure time and overheads (Table~\ref{tab:obslog}); individual runs lasted up to $\sim$6.0\,h, and we use the longest sequence as our primary dataset for the detailed PSD analysis, since its long, nearly continuous coverage provides the best frequency resolution and signal-to-noise ratio, whereas the shorter runs mainly serve as consistency checks.

\subsection{SOAR/Goodman Spectroscopy}
\label{subsec:soar}

\textbf{We obtained a SOAR/Goodman long-slit spectrum of EP240309a with a 450\,s exposure, using the 400\_SYZY grating, a $1\farcs0$ slit, and $2\times2$ on-chip binning, at an airmass of $\simeq 1.24$.} \textbf{The data were reduced with the \textsc{PypeIt} Python package \citep{pypeit:joss_pub,pypeit:zenodo}, following its standard long-slit workflow, including standard detector-level processing and overscan correction, and were subsequently flux-calibrated with the spectrophotometric standard EG\,21 and corrected for telluric absorption.} \textbf{From the arc quality assurance products we measure a dispersion of $\simeq 2.0\,\mathrm{\AA\,pix^{-1}}$ and arc-line widths of ${\rm FWHM}_{\lambda} \approx 6\,\mathrm{\AA}$, implying a resolving power of $R\sim700$--$1100$ across our wavelength range, with higher $R$ toward the red.}

\subsection{\textit{TESS} Data Products and Light-curve Extraction}
\label{subsec:tess_data}

\textbf{For the space-based optical variability analysis, we used calibrated \textit{TESS} full-frame image (FFI) cutouts retrieved from MAST with the \textsc{TESScut} service, which provides target-pixel-style cutouts generated with \textsc{Astrocut}. We then used \textsc{Lightkurve} to perform aperture photometry on the cutout pixel data and to extract light curves at the native FFI cadence of each sector. Cadences flagged as poor quality in the FFI products and measurements with non-finite flux values were removed. The same extraction aperture and filtering procedure were applied to all sectors to ensure a uniform data set for the subsequent timing and rms--flux analyses. Times are given in BTJD (BJD$_{\rm TDB}-2457000$).}


\section{Methods}\label{sec:methods}

\subsection{Light-curve Preparation}\label{subsec:lc_prep}

\textbf{For the stochastic-variability analysis we start from the calibrated BOOTES light curves described in Section~\ref{sec:data}. We remove non-detections, obvious outliers (e.g., cosmic-ray events), and duplicated timestamps. Unless stated otherwise, each light curve is median-normalized and mean-subtracted before computing PSDs and rms--flux statistics.}

\textbf{To suppress leakage from known coherent modulations into the broadband continuum, we prewhiten the spin, beat, and orbital fundamentals and their first harmonics using fixed published frequencies. We emphasize that this step targets only narrow line components and does not impose or enhance any continuum break.}

\textbf{Specifically, we fit and subtract a fixed-frequency multi-sinusoid model}
\begin{equation}\label{eq:prewhiten}
x(t)=c_{0}+\sum_{j\in\{\rm spin,beat,orb\}}\sum_{h=1}^{2}
\left[a_{j,h}\cos\!\left(2\pi h f_j t\right)+b_{j,h}\sin\!\left(2\pi h f_j t\right)\right],
\end{equation}
\textbf{where $x(t)$ is the median-normalized light curve, and $f_{\rm spin}$, $f_{\rm orb}$, and $f_{\rm beat}\equiv f_{\rm spin}-f_{\rm orb}$ are fixed to reported values\citep{potter2024}. PSDs are computed from the residuals after subtracting the best-fitting model.}

\textbf{For \textit{TESS}, we retrieve FFI cutouts via MAST TESScut (\texttt{astrocut}; \citealt{Brasseur2019}) and extract simple aperture photometry light curves using \texttt{lightkurve} \citep[e.g.,][]{Lightkurve2018}. We remove NaNs and retain only cadences with \texttt{QUALITY==0}. Time stamps are in $\mathrm{BJD}_{\mathrm{TDB}}$; any constant offset applied for plotting is purely cosmetic and does not affect the analysis. EP240309a is covered in Sectors~10, 37, 64, 90, 99, and 100.}

\textbf{Because BOOTES nightly windows are short and fragmented, ground-based period searches are strongly affected by aliases and red-noise leakage (Section~\ref{subsec:periodicity_checks}). Throughout, we therefore adopt the published \textit{TESS}-based spin and orbital periods for prewhitening and interpretation.}



\subsection{Power Spectral Density Analysis}\label{subsec:methods_psd}

\textbf{We compute PSDs in fractional rms$^{2}$\,Hz$^{-1}$ units using Welch's method with 50\% overlap \citep{Welch1967}. Each cleaned light curve is interpolated onto a uniform grid with cadence $\Delta t$ (median positive time step), then median-normalized and mean-removed. Continuum fits are performed over a predefined frequency band (given in figure captions), typically $f_{\rm low}\simeq 1/(3T_{\rm span})$ and $f_{\rm high}\lesssim 0.5f_{\rm Nyq}$ with $f_{\rm Nyq}=1/(2\Delta t)$.
We fit two alternative continuum models in $\log_{10}P$--$\log_{10}f$ space: a single power law (SPL) and a bending power law (BPL)},
\begin{equation}
P_{\rm SPL}(f)=Af^{-\alpha}, \qquad
P_{\rm BPL}(f)=A\left(\frac{f}{f_{\rm b}}\right)^{-\alpha_{\rm low}}
\left[1+\left(\frac{f}{f_{\rm b}}\right)^{(\alpha_{\rm high}-\alpha_{\rm low})}\right]^{-1}.
\end{equation}



\subsection{Model Selection and One-sided Constraints}\label{subsec:modelsel}
We compare SPL and BPL fits using AIC/BIC (least squares in $\log_{10}P$ with a Gaussian likelihood), defining $\Delta{\rm IC}\equiv {\rm IC}({\rm SPL})-{\rm IC}({\rm BPL})$. To guard against spurious bends from red noise, we test the SPL null with Monte Carlo surrogates generated following \citet{Timmer1995}. We claim a bend only when information criteria and the red-noise test consistently favor the BPL; otherwise we report one-sided lower limits on $f_{\rm b}$ from injection--recovery simulations, which translate (under $f_{\rm b}\sim f_{\rm K}(R_{\rm in})$) into conservative upper bounds on a characteristic inner radius \citep[e.g.,][]{Revnivtsev2009,Scaringi2012}.



\subsection{TESS Rms--flux Relation}\label{subsec:rmsflux_method}
For each \textit{TESS} sector we compute rms--flux relations using non-overlapping 0.75/1.0/1.5\,h windows. The mean flux is measured from the median-normalized light curve; the rms is computed from residuals after subtracting a linear trend within each window. We fit $\mathrm{RMS}=a\langle F\rangle+b$ and estimate slope uncertainties via bootstrap resampling (median and 16th--84th percentiles) \citep[e.g.,][]{uttley2005,Scaringi2012}.


\subsection{Spectroscopy: Deconvolved Line Widths and Radius Mapping}\label{subsec:spec_radius}

\textbf{Instrumental-broadening corrections are applied in quadrature but are negligible for the very broad Balmer and He\,\textsc{ii} lines considered here and do not affect our results.}
\textbf{To express the observed widths as a characteristic velocity scale, we adopt a fiducial white-dwarf mass $M_{\rm WD}=0.8\,M_\odot$ \citep[e.g.,][]{Pala2022} and a representative inclination $i=60^\circ$ (with a systematic envelope for $i\in[45^\circ,75^\circ]$).}
We convert between mass and radius using the zero-temperature mass--radius relation of \citet{Nauenberg1972}.
If one formally interprets a deconvolved velocity width $v_{\rm FWHM}$ as tracing rotation, a characteristic Keplerian radius is
\begin{equation}\label{eq:r_from_fwhm}
r \simeq \frac{G M_{\rm WD}}{\big[\tfrac{v_{\rm FWHM}}{2\sin i}\big]^2}.
\end{equation}

\textbf{However, as discussed in Section~\ref{Spectroscopy}, H$\beta$ can be multi-component and phase-dependent, so the optical line profiles need not arise purely from a steady Keplerian disc.}
\textbf{Because our single-epoch profiles are broadly single-peaked (preventing a peak-separation measurement), any mapping from $v_{\rm FWHM}$ to $r$ is model-dependent and should be regarded only as an order-of-magnitude characteristic scale, not a measurement of a disc radius.}
\textbf{Non-disc broadening (e.g., stream/curtain/wind or turbulence) would tend to increase $v_{\rm FWHM}$ and thus bias the formal $r$ low; accordingly, we do not use Eq.~(\ref{eq:r_from_fwhm}) to claim disc-fed accretion or to place quantitative constraints on the magnetospheric truncation radius $R_{\rm m}$.}


\section{Results}
\label{sec:results}

\subsection{Light Curves}

\begin{deluxetable*}{l l l l r r r r}
\tabletypesize{\scriptsize}
\setlength{\tabcolsep}{4.0pt}
\tablewidth{\textwidth}

\tablecaption{Log of BOOTES observations (UTC time range, filter, exposure time, and number of photometric points used in Fig.~\ref{fig:individual_curves}). Times are in UTC; $t_0$ is the start time of each run.\label{tab:obslog}}

\tablehead{
\colhead{Telescope} &
\colhead{Start (UTC)} &
\colhead{End (UTC)} &
\colhead{Filt.} &
\colhead{Exp. (s)} &
\colhead{$N$} &
\colhead{Dur. (h)} &
\colhead{Cad. (s)}
}
\startdata
BOOTES-6 & 2024-03-21 22:47:28 & 2024-03-21 23:34:43 & Clear & 180 & 11  & 0.7876 & 335.1 \\
BOOTES-6 & 2024-03-22 21:01:15 & 2024-03-23 02:59:06 & Clear & 30  & 114 & 5.9642 & 64.4  \\
BOOTES-7 & 2024-03-22 08:39:43 & 2024-03-22 09:59:35 & Clear & 30  & 77  & 1.3311 & 48.0  \\
BOOTES-7 & 2024-03-23 01:01:10 & 2024-03-23 06:28:33 & Clear & 30  & 267 & 5.4564 & 52.0  \\
BOOTES-6 & 2024-03-24 19:27:15 & 2024-03-24 23:59:56 & Clear & 30  & 214 & 4.5446 & 63.8  \\
BOOTES-7 & 2024-03-25 01:11:15 & 2024-03-25 07:07:06 & Clear & 30  & 318 & 5.9308 & 54.0  \\
\enddata
\end{deluxetable*}

\begin{figure}[htbp]
    \centering
    \includegraphics[width=\columnwidth]{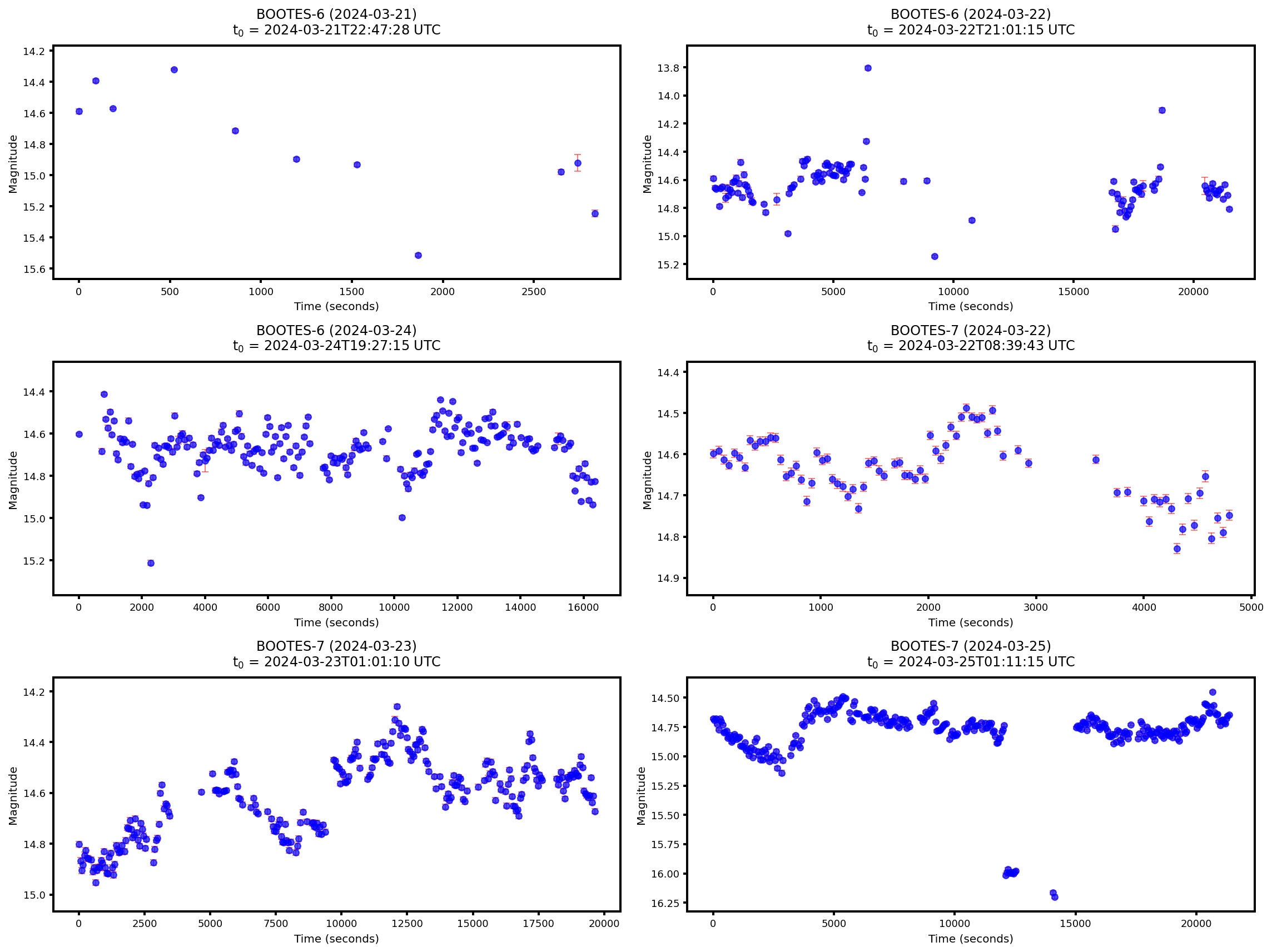}
    \caption{
    BOOTES-6 and BOOTES-7 optical light curves of EP240309a from 2024 March 21 to 25.
    Each panel shows one night of differential photometry, with time measured from the start of that night's observations on the horizontal axis and magnitude on the vertical axis.
    The corresponding reference time $t_0$ (i.e., $t=0$) is indicated in each panel in UTC.
    Blue points indicate individual measurements and red bars show the total photometric uncertainty (Section~\ref{sec:data}).
    All nights display pronounced variability on minute-to-minute timescales
    on top of modest night-to-night changes in mean brightness.
    }
    \label{fig:individual_curves}
\end{figure}

The BOOTES-6 and BOOTES-7 light curves (Fig.~\ref{fig:individual_curves}) provide six nights of coverage over the interval 2024 March 21--25, with individual continuous segments ranging from $\sim0.7$ to $\sim6.0$~h (Table~\ref{tab:obslog}).
The bulk of the measurements fall in the range $14.3 \lesssim R \lesssim 15.6$\,mag
(Fig.~\ref{fig:individual_curves}), with a few isolated points outside this interval; night-averaged levels differ by at most a few tenths of a magnitude, \textbf{while all nights show strong variability on timescales of minutes and longer, i.e., broadband flickering commonly seen in accreting CVs. Such variability is not unique to disc accretion and can also arise in stream-fed or curtain-dominated geometries, so the BOOTES light curves alone do not discriminate the accretion mode.}


\subsection{Periodicity Checks}
\label{subsec:periodicity_checks}

\textbf{Because the nightly windows are short and the light curves are dominated by red-noise flickering, period searches in the combined, cleaned \textit{BOOTES} dataset (913 points over a time baseline of 58.1\,h) recover only low significant candidate features rather than independently significant periodicities.} Using standard LS and PDM analyses over 0.5--5.0\,h, we find an LS peak at 2.0300\,h and a PDM minimum at 3.5345\,h, but red-noise surrogate tests give FAP$>0.05$ for both. The 2.0300\,h LS peak is consistent with the 24\,h sampling alias of the previously reported harmonic at $P_{\rm orb}/2=1.8807$\,h, and the 3.5345\,h feature is likewise compatible with a daily-window alias. We therefore interpret these signals as low-significance aliases rather than true detections. The BOOTES data are thus consistent with the published \textit{TESS}-based orbital and spin periods \citep{potter2024,xiao2025}, but do not provide additional timing constraints, so we adopt the published periods for narrow-line prewhitening prior to continuum PSD modeling.


\subsection{Power Spectral Density}\label{sec:results_psd}

\begin{figure*}[htbp]
  \centering
  \includegraphics[width=\textwidth]{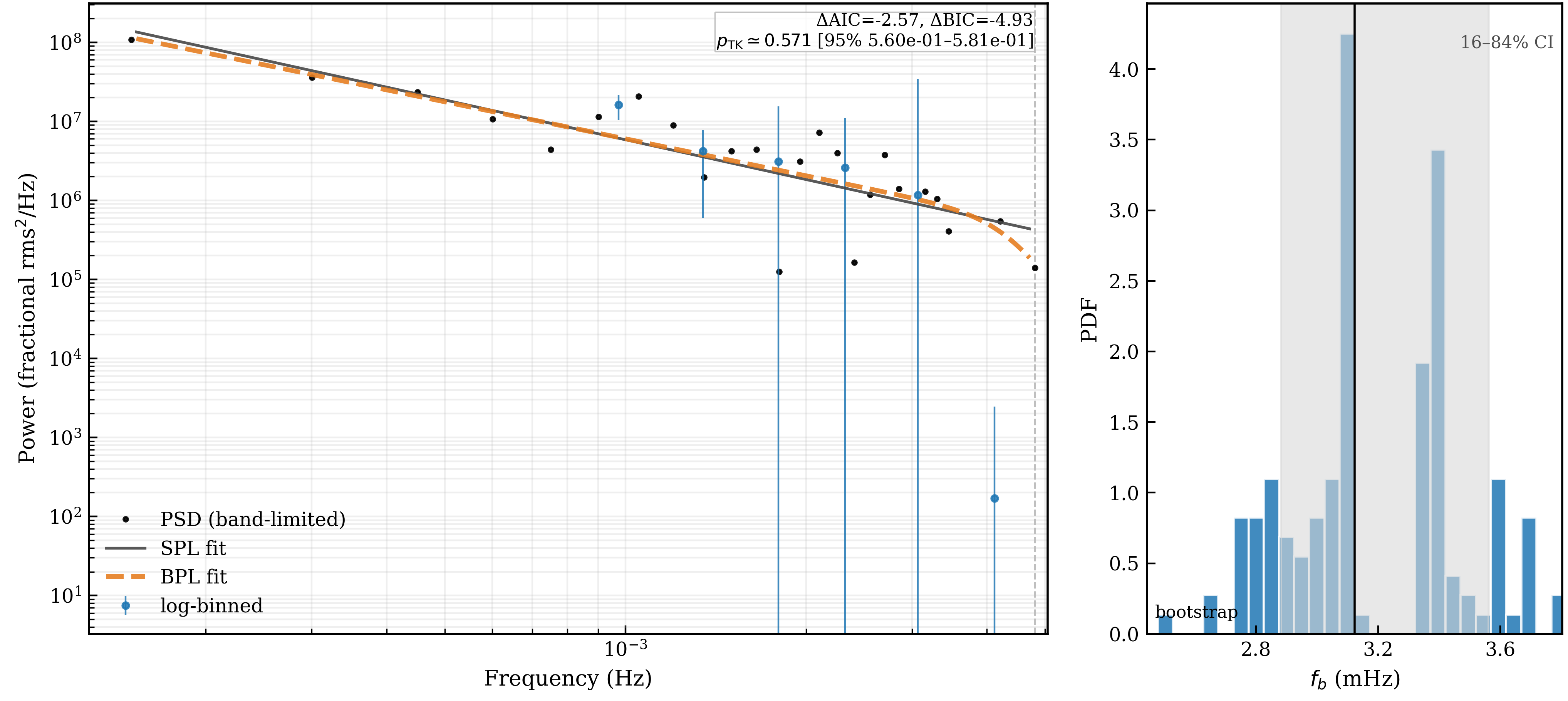}
  \caption{\textbf{Full-night BOOTES PSD (fractional rms$^{2}$\,Hz$^{-1}$) for the longest sequence, with the adopted analysis band and best-fitting SPL/BPL continua overplotted. The SPL continuum is mildly preferred over the BPL and TK simulations do not reject the SPL null hypothesis, so no bend is claimed within the sampled frequency range.}}
  \label{fig:psd_full}
\end{figure*}

\textbf{Figure~\ref{fig:psd_full} shows the PSD of the longest BOOTES sequence and the best-fitting continuum models. The SPL continuum is mildly preferred over the BPL ($\Delta$AIC $=-2.57$, $\Delta$BIC $=-4.93$), and TK red-noise simulations matched to the best-fitting SPL do not reject the SPL null ($p_{\rm TK}\simeq0.57$). We therefore do not claim evidence for a bend in the sampled frequency range.}

\textbf{We convert this non-detection into one-sided lower limits on the bend frequency $f_{\rm b}$ from injection--recovery simulations (fiducial strict convention):}
\[
\mathrm{UL}_{95\%}=0.395~\mathrm{mHz},\qquad
\mathrm{UL}_{90\%}=0.334~\mathrm{mHz}.
\]
\textbf{In the remainder of the paper we adopt $\mathrm{UL}_{90\%}$ as the fiducial constraint, implying conservative upper bounds on a characteristic inner scale; if one adopts the standard dynamical identifications, these can be expressed as upper bounds on $R_{\rm m}$ under the adopted mappings. For $M_{\rm WD}=0.8\,M_\odot$, we obtain $R_{\rm m}/R_{\rm WD}\lesssim41.4$ (dynamical) and $\lesssim30.6$ (free-fall), adopting $R_{\rm WD}\simeq7.0\times10^{8}$\,cm from the zero-temperature mass--radius relation of \citet{Nauenberg1972}.}

\textbf{As a consistency check, SPL fits to individual BOOTES nights within the common cross-night analysis band yield mutually consistent slopes and are not used for physical inference.}

\subsection{TESS Rms--flux Relation}\label{subsec:rmsflux}

\textbf{We test for a linear rms--flux relation on hour timescales \citep[e.g.,][]{uttley2005,Scaringi2012}. Using the same extraction and filtering across sectors, we find clear positive linear trends in Sectors~37, 64, and 99, whereas Sectors~90 and 100 do not show a robust detection; Sector~10 provides no valid high-cadence segments at these window sizes and is excluded.}

\textbf{For the fiducial 1\,h windows, the Pearson coefficients are $r_{\rm P}=0.336$ ($p_{\rm P}=1.3\times10^{-16}$; $N_{\rm seg}=572$) in Sector~37, $r_{\rm P}=0.195$ ($p_{\rm P}=9.2\times10^{-7}$; $N_{\rm seg}=624$) in Sector~64, $r_{\rm P}=0.271$ ($p_{\rm P}=2.4\times10^{-9}$; $N_{\rm seg}=470$) in Sector~99, $r_{\rm P}=-0.0158$ ($p_{\rm P}=0.689$; $N_{\rm seg}=642$) in Sector~90, and $r_{\rm P}=0.0302$ ($p_{\rm P}=0.46$; $N_{\rm seg}=609$) in Sector~100. We treat $r_{\rm P}$ as descriptive; the primary effect size is the bootstrapped linear slope.}

\textbf{At 1\,h, the bootstrap-median slopes are $a_{50}=4.34\times10^{-4}$ for Sector~37, $2.55\times10^{-3}$ for Sector~64, and $3.93\times10^{-3}$ for Sector~99, with confidence intervals that remain positive, while Sector~100 gives $a_{50}=1.95\times10^{-2}$ but with an interval spanning zero. The positive trend also persists across 0.75/1.0/1.5\,h windows in Sectors~37, 64, and 99, whereas Sectors~90 and 100 remain state-dependent non-detections in our dataset.}

\textbf{Using the same fixed-aperture extraction and QUALITY=0 filtering, the median aperture-summed flux levels are $\tilde{F}_{37}=9572$, $\tilde{F}_{64}=9786$, and $\tilde{F}_{99}=9584$, whereas Sectors~90 and 100 are both consistent with zero in the same extracted units ($\tilde{F}_{90}=-1.33$, $\tilde{F}_{100}=0.051$). We therefore do not interpret Sectors~90/100 as straightforward absolute-brightness comparisons to Sectors~37/64/99; rather, they indicate that the rms--flux relation is not robustly detected in all epochs with the same pipeline.}


\subsection{Optical Spectroscopy}
\label{Spectroscopy}

\begin{figure}[htbp]
  \centering
  \includegraphics[width=\columnwidth]{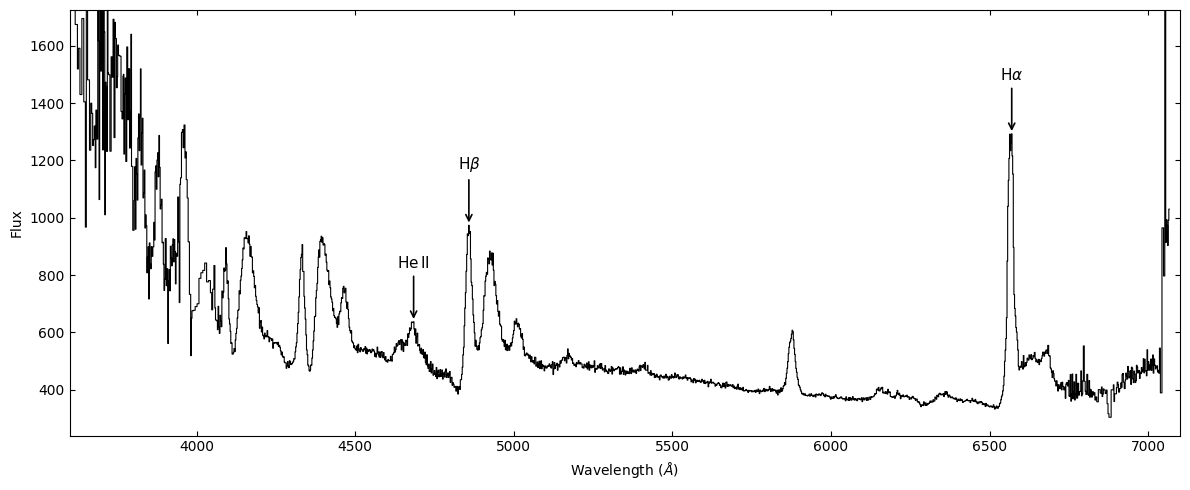}
  \caption{\textbf{SOAR/Goodman optical spectrum of EP240309a at native sampling/resolution.
  Major emission features (H$\alpha$, H$\beta$, He\,{\sc ii}~4686) are marked.
  Broad low-contrast structures near $\sim4160$, $\sim4400$, and $\sim4940$~\AA\ are indicated and discussed in the text.}}
  \label{fig:optical_spectrum}
\end{figure}

\textbf{We obtained flux-calibrated SOAR/Goodman long-slit spectroscopy of EP240309a on 2024-03-25 (start 06:59~UTC; Fig.~\ref{fig:optical_spectrum}) to document the spectral state contemporaneous with our March 2024 variability campaign.
\citet{potter2024} present SALT/RSS spectroscopy (low resolution, $R\sim800$) obtained on 2024-03-19 with orbital-phase-resolved analysis and show that H$\beta$ comprises multiple components; therefore, the line profiles cannot be interpreted as arising purely from a steady Keplerian accretion disc.
Here we use our single-epoch spectrum only for basic line-profile characterization and descriptive (velocity-scale) checks.}

\textbf{The wavelength solution and arc-line widths imply a characteristic instrumental FWHM of $\Delta\lambda \simeq 6.2$~\AA\ (median arc FWHM $\simeq 3.11$~pix and dispersion $\simeq 2.0$~\AA~pix$^{-1}$), corresponding to $R\sim670$ at 4160~\AA, $R\sim707$ at 4400~\AA, $R\sim805$ at 5000~\AA, and $R\sim1070$ near H$\alpha$.
In addition to the prominent Balmer and He\,{\sc ii} emission, Fig.~\ref{fig:optical_spectrum} shows broad residual structures (bumps) around $\sim4160$, $\sim4400$, and $\sim4940$~\AA.
These low-contrast broad bumps are not apparent in the 2024-03-19 SALT/RSS spectrum \citep{potter2024}, plausibly because they are state-dependent and can be diluted or washed out by differences in the continuum level, S/N, and spectral resolution and/or continuum normalization.
To quantify these features in a reproducible manner, we fit a local linear pseudo-continuum in sidebands away from strong lines and measure the residual excess in a core window using the telluric-corrected 1D spectrum.
For the three features, the residual peaks at $\approx(48\%,78\%,89\%)$ above the local pseudo-continuum with peak ${\rm S/N}\approx(31,48,62)$, and remains highly significant when integrated over the core windows (${ \rm S/N}\approx(122,210,258)$; $EW_{\rm pos}\approx(19.8,36.4,38.7)$~\AA, respectively).
Given the single-epoch nature and modest resolution of our spectrum, we do not attempt a unique identification of these weak broad structures here, but we report their presence and note that their contrast can plausibly vary with accretion state and continuum placement.}

\begin{figure}[htbp]
  \centering
  \includegraphics[width=\columnwidth]{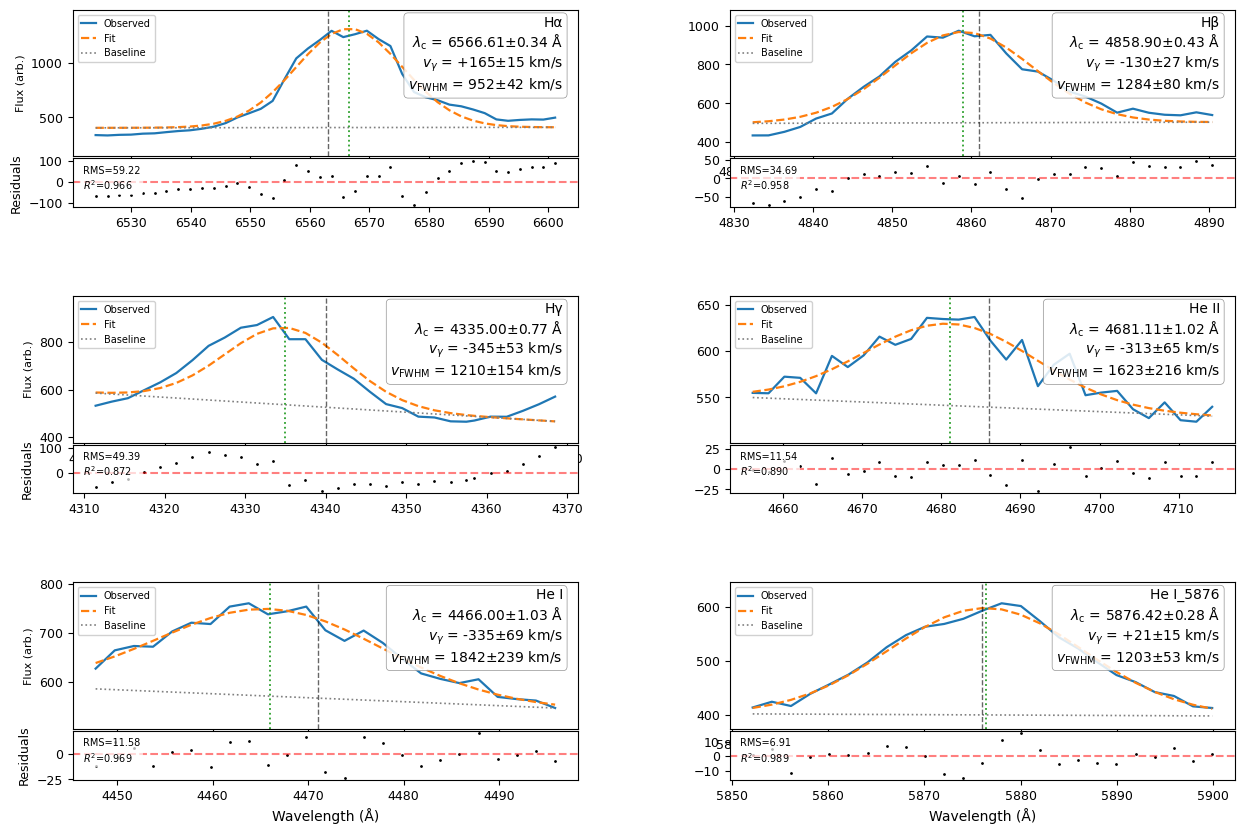}
  \caption{Example Gaussian$+$linear-baseline fits to six prominent emission lines used to estimate centroids and deconvolved widths.
  We annotate the fitted centroid $\lambda_{\rm c}$ and the corresponding apparent velocity offset $v_\gamma=c(\lambda_{\rm c}/\lambda_{0}-1)$ in each panel; the annotated $v_{\rm FWHM}$ values are derived from deconvolved widths (see Section~\ref{subsec:spec_radius}).}
  \label{fig:emission}
\end{figure}

\textbf{Deconvolved velocity widths cluster around $v_{\rm FWHM}\sim(1$--$2)\times10^{3}$~km\,s$^{-1}$ (Fig.~\ref{fig:emission}).
The fitted centroid offsets $v_\gamma$ are not consistent in sign across wavelength, indicating that line-by-line systematics (e.g., profile asymmetries/blends and residual wavelength-calibration zero-point errors) dominate over any single bulk velocity in this single-epoch spectrum.
Balmer profiles show enhanced broad wings, most evident in H$\beta$ and H$\alpha$.
To assess whether such wings bias our width measurements, we repeated the local fits allowing an additional broad component; the inferred core FWHM changes by $\lesssim 12\%$ (H$\beta$: $-3.9\%$; H$\alpha$: $-11.5\%$) relative to single-Gaussian fits, i.e., within the systematic uncertainties of our line-profile characterization.
The observed profiles are broadly single-peaked with mild red-wing asymmetries, but any phase-dependent substructure would be smeared in this single-epoch spectrum.
Accordingly, any line-width-based radius mapping should be interpreted only as an order-of-magnitude characteristic scale rather than as a precise geometric constraint.}


\section{Discussion}
\label{sec:discussion}

\subsection{Rms--flux Behavior and Stochastic Variability}

\textbf{Our \textit{TESS} analysis shows that the linear rms--flux relation in EP240309a is epoch-dependent. Positive hour-timescale trends are detected in Sectors~37, 64, and 99, whereas Sectors~90 and 100 do not show a robust detection, and Sector~10 provides no valid high-cadence segments at the adopted window sizes (Section~\ref{subsec:rmsflux}). We therefore interpret the non-detections as state-dependent absences of a measurable trend in our dataset, rather than as external control cases. Using the same fixed-aperture extraction and QUALITY=0 filtering, Sectors~37/64/99 have similar positive median aperture-summed flux levels ($\tilde{F}_{37}=9572$, $\tilde{F}_{64}=9786$, $\tilde{F}_{99}=9584$), whereas Sectors~90/100 are consistent with zero in the same extracted units ($\tilde{F}_{90}=-1.33$, $\tilde{F}_{100}=0.051$), so we do not interpret these sectors as straightforward absolute-brightness comparisons.}

\textbf{The detections in Sectors~37, 64, and 99 are consistent with multiplicative variability processes widely observed in accreting compact objects \citep[e.g.,][]{uttley2005,Scaringi2012}. For the present data quality, a simple linear description is adequate; more flexible forms are not required by the information criteria and do not change the qualitative behavior. We therefore use the bootstrapped slope as the primary effect size and treat the Pearson statistics listed in Section~\ref{subsec:rmsflux} as descriptive diagnostics.}

\textbf{The rapid optical variability and PSD constraints further indicate broadband red-noise behavior typical of cataclysmic variables: the fitted single--power--law indices span $\alpha\simeq 1.16\pm0.23$ to $1.78\pm0.11$ across nights (Section~\ref{sec:results_psd}), consistent with stochastic accretion-driven flickering \citep[e.g.,][]{Lyubarskii1997,uttley2005,Bruch2021}. However, we do not detect a robust PSD bend in the sampled frequency range, so the timing analysis provides only one-sided lower limits on any putative bend frequency. Under standard dynamical identifications, these translate into conservative upper bounds on a characteristic inner scale and do not constitute evidence for a measured magnetospheric truncation radius or for disc-fed accretion in EP240309a.}

\textbf{Finally, the contemporaneous optical spectroscopy (Section~\ref{Spectroscopy}) also cautions against over-interpreting the geometry: the available single-epoch, coadded spectrum does not provide phase-resolved kinematics, and published time-resolved spectra indicate that at least H$\beta$ contains multiple components. Any radius estimate based on deconvolved line widths should therefore be treated only as an order-of-magnitude check rather than as a quantitative constraint on the magnetospheric truncation radius.}


\subsection{Accretion-flow Geometry and the Magnetospheric Radius}
\label{subsec:geometry_rm_discussion}

\textbf{EP240309a has $P_{\rm orb}=3.762$\,h and $P_{\rm spin}=3.97$\,min, implying $P_{\rm spin}/P_{\rm orb}\approx0.018$, within the range of IP candidates \citep[e.g.,][]{patterson1994,warner1995,mukai2017}. However, this timing ratio alone does not establish the detailed accretion geometry, so we treat EP240309a as an IP candidate.}

\textbf{Because we do not detect a robust PSD bend (Section~\ref{sec:results_psd}), our timing analysis provides only one-sided constraints on a characteristic inner scale; under standard dynamical identifications, these may be expressed as conservative upper bounds on $R_{\rm m}$. We therefore regard them as broad consistency checks rather than as evidence for a measured magnetospheric truncation radius.}

\textbf{For reference, the corotation radius is}
\[
R_{\rm co}=\left(\frac{GM_{\rm WD}P_{\rm spin}^{2}}{4\pi^{2}}\right)^{1/3}.
\]

\textbf{Emission-line widths provide only an order-of-magnitude velocity-scale check: the available single-epoch, coadded profiles lack phase-resolved kinematics, and published time-resolved spectroscopy shows multi-component H$\beta$ structure.}

\textbf{Overall, the present data are consistent with accretion onto a magnetic white dwarf, but do not allow us to establish the detailed accretion geometry; in particular, they do not demonstrate disc truncation or distinguish robustly between disc-fed and stream-/curtain-dominated scenarios. Longer, less-windowed optical baselines together with contemporaneous time-resolved spectroscopy are required for stronger geometric constraints.}


\subsection{Transient Behavior and Implications for the IP Population}

\textbf{EP240309a was discovered as an X-ray transient by \textit{Einstein Probe}; during our March 2024 monitoring its optical brightness spans $\Delta R\simeq1.3$\,mag, but our BOOTES/\textit{TESS} coverage does not capture a full rise/decay cycle and therefore cannot establish a classical dwarf-nova--like optical outburst evolution. In line with this, the ASAS-SN Sky Patrol photometry reported by \citet{potter2024} is more consistent with a bimodal brightness distribution (i.e., state-dependent switching) than with well-sampled, discrete outburst rise--decay cycles.}

\textbf{Several mechanisms could drive transient episodes in magnetic cataclysmic variables (e.g., disc instabilities or gated accretion; \citealt{lasota2001,DAngeloSpruit2010,Scaringi2017,Dubus2024}), but the present data cannot distinguish among them. The observed flickering and epoch-dependent rms--flux behavior indicate stochastic variability, but not a specific trigger mechanism; discriminating among scenarios requires simultaneous X-ray/optical coverage through full transient cycles.} \textbf{EP240309a adds to the emerging sample of transient IP-like systems by showing red-noise optical variability and epoch-dependent rms--flux behavior consistent with accretion onto a magnetic white dwarf, while its detailed accretion geometry and transient trigger remain unconstrained. Longer multi-wavelength monitoring is needed to assess recurrence and to place the source more robustly within the IP population.}


\section{Conclusions}\label{sec:conclusions}

\textbf{We present multi-night BOOTES optical photometry and a contemporaneous SOAR/Goodman spectrum of EP240309a, complemented by archival \textit{TESS} photometry, to characterize its stochastic optical variability and to test for an rms--flux relation. The present data yield only conservative, one-sided constraints on a characteristic inner scale and do not establish the detailed accretion geometry; accordingly, we describe EP240309a as an intermediate-polar candidate rather than a confirmed transient intermediate polar.}

\textbf{The BOOTES PSD of the best night is well described by a single power law and shows no compelling evidence for a bend in the sampled frequency range. Injection--recovery simulations yield strict one-sided lower limits on any bend frequency, which translate into order-of-magnitude constraints on a characteristic inner scale under standard dynamical identifications and subject to unknown system parameters}. For $M_{\rm WD}=0.8\,M_\odot$, these limits can be expressed (under standard dynamical identifications) as $R_{\rm m}/R_{\rm WD}\lesssim 41.4$ (dynamical) or $\lesssim 30.6$ (free-fall).

\textbf{Space-based data reveal a linear rms--flux relation on hour timescales in three high-cadence \textit{TESS} sectors (37, 64, and 99), while Sectors~90 and 100 do not show a robust detection using the same analysis pipeline, indicating that the rms--flux behavior is epoch-dependent in EP240309a; Sector~10 is unusable for this test because no valid high-cadence segments are available at the adopted window sizes. For our fiducial 1\,h windows we obtain slopes $a=(4.43^{+0.86}_{-0.82})\times10^{-4}$ (Sector~37), $a=(2.55^{+0.62}_{-0.68})\times10^{-3}$ (Sector~64), and $a=(3.93^{+0.99}_{-0.94})\times10^{-3}$ (Sector~99), in units of fractional rms per unit mean flux. Together with the red-noise PSDs, these results are consistent with multiplicative variability processes, but do not by themselves identify the underlying accretion geometry.}

The SOAR spectrum shows Balmer and He\,\textsc{ii} emission with ${\rm FWHM}\approx1000$--$1600$\,km\,s$^{-1}$, with He\,\textsc{ii}~$\lambda4686$ systematically broader than the Balmer lines. \textbf{Given the inclination dependence of line-width-based radius estimates and published evidence for multi-component Balmer-line structure, spectroscopy provides at most an order-of-magnitude cross-check rather than a quantitative geometric constraint.} Future longer, contiguous high-cadence monitoring and coordinated X-ray/optical observations will be key to tightening inner-scale constraints and testing for state changes.


\begin{acknowledgments}
This work was supported by the China Scholarship Council (CSC). We acknowledge the use of data from the BOOTES (Burst Observer and Optical Transient Exploring System) network. This research is based in part on observations obtained at the Southern Astrophysical Research (SOAR) telescope. We thank the Instituto de Astrofísica de Andalucía (IAA-CSIC) for the support and collaboration in this research. AJCT acknowledges the Spanish Ministry of Science, Innovation and Universities project PID2023-151905OB-I00, and the Centro de Excelencia Severo Ochoa grant CEX2021-001131-S funded by MCIN/AEI/10.13039/501100011033. The program of development within Priority-2030 is acknowledged for supporting the research at UrFU (04.89). \textbf{MCG acknowledges financial support from the Spanish Ministry project MCI/AEI/PID2023-149817OB-C31 and the Severo Ochoa grant CEX2021-001131-S funded by MICIU/AEI/10.13039/501100011033.}

\end{acknowledgments}

\facilities{SOAR (Goodman) \citep{Clemens2004SPIE},
BOOTES-6, BOOTES-7 \citep{Castro1999,hu2023,CastroTirado2023NatAstron},
TESS \citep{Ricker2015}}

\software{Astropy \citep{Astropy2013,Astropy2018,Astropy2022},
NumPy \citep{NumPy2020},
SciPy \citep{SciPy2020},
Matplotlib \citep{Matplotlib2007},
Lightkurve \citep{Lightkurve2018},
Astrocut \citep{Brasseur2019},
PypeIt \citep{pypeit:joss_pub,pypeit:zenodo}
}

\section*{Data Availability}
The BOOTES photometry is provided as machine-readable data behind Figure~\ref{fig:individual_curves} (CSV format), together with a short README describing the columns and time system.
\textbf{The \textit{TESS} time-series pixel cutouts used in this work were obtained from the calibrated full-frame images (FFIs) via MAST.
The underlying calibrated FFIs are available at MAST for Sectors~10 \citep{tess_sector10_2024},
37 \citep{tess_sector37_2024},
64 \citep{tess_sector64_2024},
90 \citep{tess_sector90_2025},
99 \citep{tess_sector99_2026}, and
100 \citep{tess_sector100_2026}.}

\clearpage
\appendix
\restartappendixnumbering

\section{Method-sensitivity Summaries}\label{app:sensitivity}

\textbf{For brevity, we summarize representative sensitivity tests for the \textit{TESS} rms--flux analysis as compact tables of the bootstrap-median slope $a_{50}$ under different quality-control settings. We include one robust positive-detection case and one non-robust-detection case at the fiducial 1\,h windows.}

\begin{table}[htbp]
\centering
\caption{Sensitivity summary for the \textit{TESS} rms--flux slope in Sector~64 (1\,h windows). The table lists the bootstrap-median linear rms--flux slope $a_{50}$ (in fractional rms per unit mean flux) obtained under different per-segment sigma-clipping thresholds and minimum-number-of-points requirements.}
\label{tab:tess_sensitivity_s64_60min}
\begin{tabular}{lccc}
\hline\hline
$\sigma$-clip (per segment) & min points $=8$ & min points $=10$ & min points $=15$ \\
\hline
3.0 & 0.00264 & 0.00260 & 0.00264 \\
2.5 & 0.00252 & 0.00252 & 0.00250 \\
2.0 & 0.00255 & 0.00251 & 0.00254 \\
\hline
\end{tabular}
\end{table}

\begin{table}[htbp]
\centering
\caption{\textbf{Sensitivity summary for the \textit{TESS} rms--flux slope in Sector~100 (1\,h windows; non-robust-detection case).} The table lists the bootstrap-median linear rms--flux slope $a_{50}$ (in fractional rms per unit mean flux) obtained under different per-segment sigma-clipping thresholds and minimum-number-of-points requirements.}
\label{tab:tess_sensitivity_s100_60min}
\begin{tabular}{lccc}
\hline\hline
$\sigma$-clip (per segment) & min points $=8$ & min points $=10$ & min points $=15$ \\
\hline
3.0 & 0.0315 & 0.0320 & 0.0324 \\
2.5 & 0.0320 & 0.0284 & 0.0358 \\
2.0 & 0.0196 & 0.0190 & 0.0226 \\
\hline
\end{tabular}
\end{table}

\bibliography{references}{}
\bibliographystyle{aasjournalv7}



\end{document}